\documentclass[%
  reprint,
  aps,
  prl,
  superscriptaddress,
  amsmath,amssymb,
  floatfix
]{revtex4-2}

\usepackage{dcolumn}
\usepackage{bm}
\usepackage[version=4]{mhchem}
\usepackage{siunitx}
\usepackage{svg}
\usepackage{graphicx,bm,mathptmx,physics,color,anyfontsize}
\usepackage[colorlinks=true,allcolors=blue]{hyperref}

\begin{document}

\title{Physical Mechanisms of Light-Induced Molecular Fragmentation in the Long-Wavelength Regime}

\author{Jesse Sandhu}
\author{Naga Krishnakanth Katturi}
\affiliation{Department of Chemistry, Michigan State University, East Lansing, MI 48824, USA}

\author{Xiao-Min Tong}
\affiliation{Center for Computational Sciences, University of Tsukuba, Tsukuba, Ibaraki 305-8573, Japan}

\author{Misha Ivanov}
\affiliation{Max Born Institute, Max-Born-Straße 2A, Berlin D-12489, Germany}
\affiliation{Department of Physics, Humboldt Universität zu Berlin, Newtonstraße 15, Berlin D-12489, Germany}
\affiliation{Technion-Israel Institute of Technology, 3200003 Haifa, Israel}

\author{Marcos Dantus}
\email{dantus@msu.edu}
\affiliation{Department of Chemistry, Michigan State University, East Lansing, MI 48824, USA}
\affiliation{Department of Physics and Astronomy, Michigan State University, East Lansing, MI 48824, USA}
\affiliation{Department of Electrical and Computer Engineering, Michigan State University, East Lansing, MI 48824, USA}

\date{\today}

\begin{abstract}
Strong near-infrared (NIR) laser fields ionize polyatomic molecules through tunneling and multiphoton processes, while subsequent molecular fragmentation can be driven by electron rescattering and field-induced nonadiabatic excitation of the molecular ion. How these competing mechanisms deposit energy into molecular ions and determine fragmentation branching remains unresolved. By combining wavelength- and ellipticity-dependent intensity-difference mass spectrometry of methanol with time-dependent Schrödinger equation (TDSE) and semiclassical Monte Carlo simulations, we demonstrate that, in the tunneling-to-intermediate regime ($\gamma \lesssim 1$), molecular fragmentation is governed by the interplay of electron recollision and field-driven nonadiabatic excitation, revealing competing physical mechanisms relevant to polyatomic molecules more broadly. Ellipticity-dependent measurements demonstrate that these mechanisms can be experimentally distinguished for individual cationic fragments. Channels requiring high internal energies exhibit the strong ellipticity-dependent suppression characteristic of rescattering-driven processes, while lower-energy channels are comparatively less sensitive. Although small-fragment yields decrease substantially with increasing wavelength, their relative distributions remain largely unchanged, indicating that the underlying fragmentation pathways are preserved and that the suppression primarily reflects reduced probabilities of electron recollision and nonadiabatic excitations at longer wavelengths.
\end{abstract}

\maketitle

Strong-field processes involving electron recollision have spurred novel dynamic imaging methods using high harmonic generation (HHG), e.g., Refs. \cite{Itatani2004OrbitalImaging,smirnova2009high,lein2002interference, lein2007molecular,uzan2020spatial,torlina2013time} and laser-induced electron diffraction, e.g. \cite{Blaga2012LIED,spanner2004reading,Lin2010LIED,schell2018molecular}. Yet, understanding strong-field dynamics in molecules remains challenging due to complex potential-energy landscapes and significant multielectron effects. 
Electron recollision, which is  central to these novel imaging approaches, can also trigger molecular fragmentation, in which impact excitation or ionization by the returning electron can initiate dissociation \cite{corkum1993MPI,Rajgara2003RescatteringFrag, Dantus_science2024}. Molecular fragmentation may also arise from inner excitation mechanisms \cite{lezius2002NME, Konar2014, nubbemeyer2008strong, mavsin2018electron}. {Here, inner excitation refers to field-driven population transfer among electronic states of the molecular ion, rather than excitation initiated by energy transfer from the returning electron. In the strong near-infrared (NIR) field, dynamic Stark shifts can bring ionic states into near resonance, enabling Landau-Zener-Dykhne-type nonadiabatic transitions \cite{Popruzhenko_2014}.}
Despite their importance, how rescattering competes with, enhances, or masks these alternative pathways in polyatomic molecules driven by the NIR field remains unclear  \cite{shchatsinin2009multiphoton}.

Wavelength is a uniquely sensitive control knob in {addressing} this problem. Fixed-energy thresholds and resonances imply that varying the photon energy provides a sensitive handle to changing the interplay of various processes.
The combination of laser wavelength, intensity, and the molecule’s ionization potential $I_p$ enter the Keldysh parameter $\gamma = \sqrt{I_p/2U_p}$, where $U_p$ is the ponderomotive energy \cite{keldysh1964ionization}. Multiphoton ionization is often assumed when $\gamma \gg1$  and tunnel ionization when $\gamma\ll1$. However, $\gamma$ is not a definitive criterion for molecular ionization because orientation-dependent potentials, multielectron effects, and nuclear dynamics violate the simplified single-active-electron 
assumptions underlying the Keldysh model \cite{keldysh1964ionization} or its molecular modifications \cite{tong2002MTI,murray2011tunnel}. 

From an electron rescattering perspective, the ponderomotive energy $U_p\propto I\lambda^2$ increases with wavelength $\lambda$ at fixed intensity $I$, raising the {maximum} return energy of 3.17$U_p$ \cite{corkum1993MPI,schafer1993above}. At the same time, increasing the wavelength allows the wavepacket to spread more during its longer travel in the continuum, decreasing the rescattering probability \cite{Niikura2003, jens2012wavelengthscaling, Villenueve2009HHG}. Disentangling  the relative roles of multiphoton {excitation}, tunneling ionization, and the mechanisms of energy deposition that activate competing fragmentation pathways is nontrivial in molecules, where resonances, intermediate states, and multielectron effects co-exist \cite{corkum1995EnhancedIonization, corkum1996Enhancedionization, Wu2012Enhancedionization, Liu2007starkshifts, Suchowski2008starkshifts}. Early strong-field mass spectrometry studies showed that irradiation at longer infrared wavelengths { $\lambda\gtrsim 1.0-1.5\mu$m} produces considerably less fragmentation compared to NIR ($\lambda=800$ nm) drivers \cite{Shusterman2022, bohinski2013measurement, lezius2001nonadiabatic, bhardwaj2004recollision}. Refs. \cite{lezius2001nonadiabatic,lezius2002NME}  explained this suppression in long linear molecules with delocalized $\pi$-electrons by attributing  fragmentation to nonadiabatic multielectron excitations, reduced at longer wavelengths. However, (i)  smaller polyatomic molecules also exhibit reduced fragmentation \cite{Tanaka2009, BOHINSKI2014}, and (ii) inelastic recollision-driven channels are also reduced at longer wavelengths \cite{corkum1993MPI, Wang2017}.

In this letter, we pair intensity-resolved time-of-flight mass spectrometry measurements with time-dependent Schr\"odinger equation (TDSE) simulations and ellipticity-dependent ion-yield measurements with semiclassical Monte Carlo simulations to assess the extent to which the inelastic electron rescattering model contributes to the strong-field-induced fragmentation of methanol as a model polyatomic system. Although the quantitative fragmentation pathways are molecule-specific, the underlying strong-field mechanisms, including recollision-driven and field-driven excitation, are not unique to methanol. Ellipticity-dependent measurements therefore provide a broadly applicable means of distinguishing competing fragmentation mechanisms in polyatomic molecules.

Anhydrous methanol was ionized over a range of intensities at three central wavelengths (800, 1300, and 1500 nm). To suppress contributions from lower-intensity regions of the focal volume, we combined a long confocal parameter and a narrow extraction slit with an intensity-difference spectroscopy (IDS) method in which ion yields measured at peak intensities $I_{0}$ and $0.9I_{0}$ were subtracted \cite{posthumus2004dynamics, Itzhak2004IDS}. Special care was taken to match the confocal parameter across all three datasets by adjusting the focus and input beam diameter to allow for a more quantitative comparison between the wavelengths. The Keldysh parameter reported in Fig. 1 was calculated using the peak intensity $I_0$.  The intensity was independently calibrated at each wavelength using the  
$\mathrm{Ar}^{2+}/\mathrm{Ar}^{+}$ ratio \cite{guo1998single}. We estimate an absolute intensity-calibration uncertainty of $\sim 30\%$ and the relative uncertainty between wavelengths at 10\%. At each wavelength $\lambda$, the intensity $I$ was adjusted to maintain a constant ponderomotive energy $U_p \propto I\lambda^2$ and thus a fixed Keldysh parameter $\gamma$ \cite{keldysh1964ionization}.
\begin{figure}[t]
  \centering
  \includegraphics[width=1\linewidth]{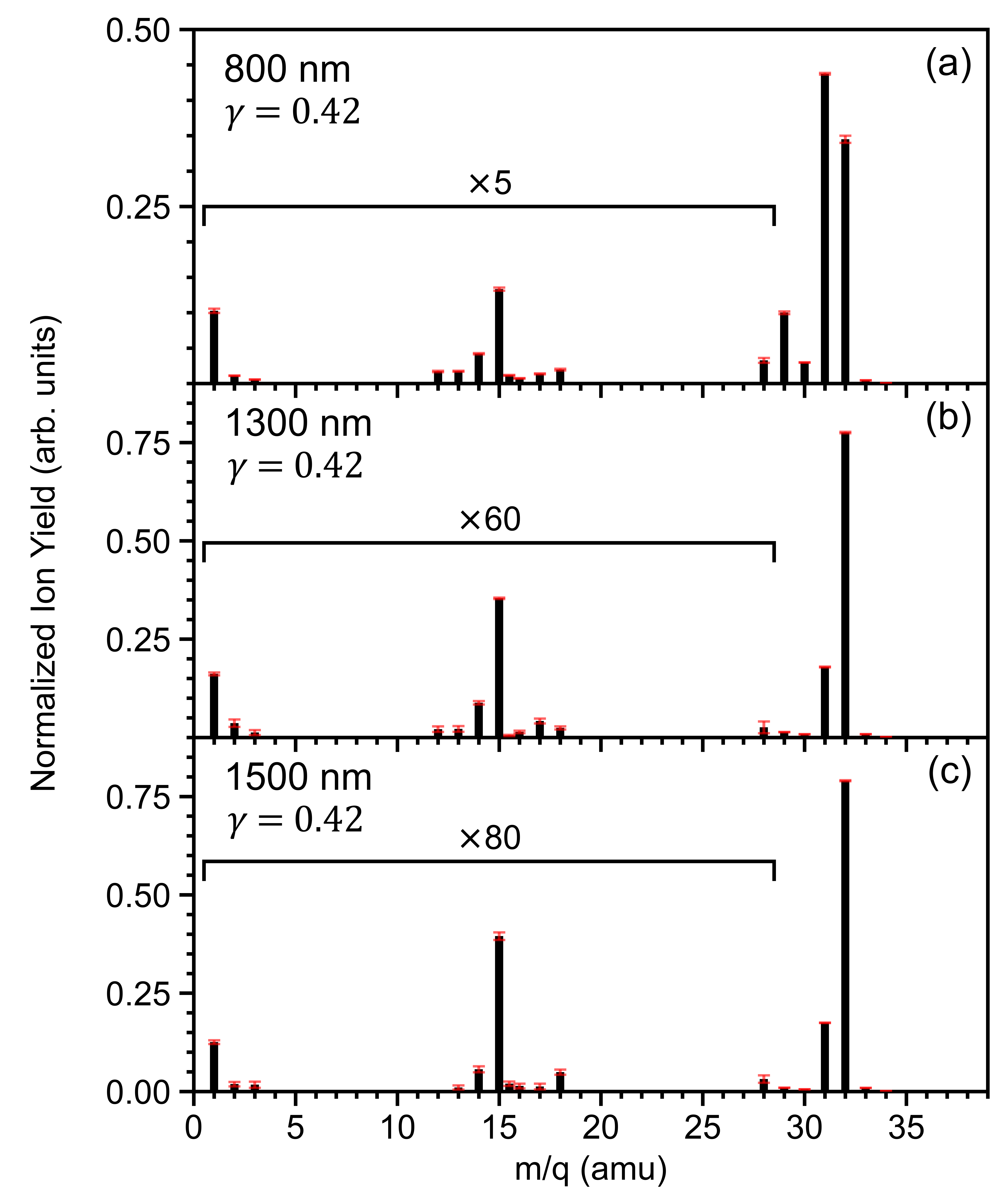}
  \caption{Mass spectra of methanol at Keldysh parameter $\gamma = 0.42$ for all three wavelengths. The IDS method ($I_0-0.9I_0$) minimizes focal-volume effects. Each spectrum is normalized independently at its respective wavelength so that the summed fragment yields equal unity. The reported order-of-magnitude reduction in small-fragment yields is determined from unnormalized measurements. Bracketed regions are magnified by the indicated factor after normalization. The red error bars indicate the standard errors of the intensity-difference ion yields. Table S1 lists the numerical values and corresponding standard errors for each m/q fragment shown in the figure.}
  \label{fig:comparison}
\end{figure}

Representative intensity-difference mass spectra in the tunneling regime ($\gamma=0.42)$ at different wavelengths for direct comparison are shown in Fig. \ref{fig:comparison}. We selected $\gamma=0.42$ to compare fragmentation across wavelengths at constant ponderomotive energy, well within the tunneling-ionization regime ($\gamma < 1)$, even allowing for the estimated absolute intensity-calibration uncertainty. The \SI{800}{nm} spectrum exhibits substantially greater fragmentation than \SI{1300}{nm} and \SI{1500}{nm}. Although this trend is expected, we find that despite more than an order-of-magnitude variation in the yields of smaller fragments (m/q $\leq 28$), the fragment identities and relative distributions remain remarkably similar across all three wavelengths after independent normalization. This suggests that the underlying fragmentation channels are largely conserved. Quantitatively, the normalized IDS distributions for the selected $m/q\leq28$ channels yield unweighted spectral entropy similarities \cite{Li2021} of 0.972, 0.941, and 0.960 for the 800–1300, 800–1500, and 1300–1500 nm comparisons, respectively, where unity indicates identical distributions.

We attribute the observed wavelength dependence in fragmentation primarily to the reduced probability of electron rescattering at longer wavelengths, while also recognizing that changes in nonadiabatic excitation pathways {and their efficiency} may contribute. For the recolliding electron wavepacket, 
a longer excursion time with increasing wavelength leads to greater spreading and reduced overlap with the parent ion, thereby lowering the rescattering probability \cite{Niikura2003}. 

\begin{figure*}[t]
  \centering
  \includegraphics[width=1.0\textwidth]{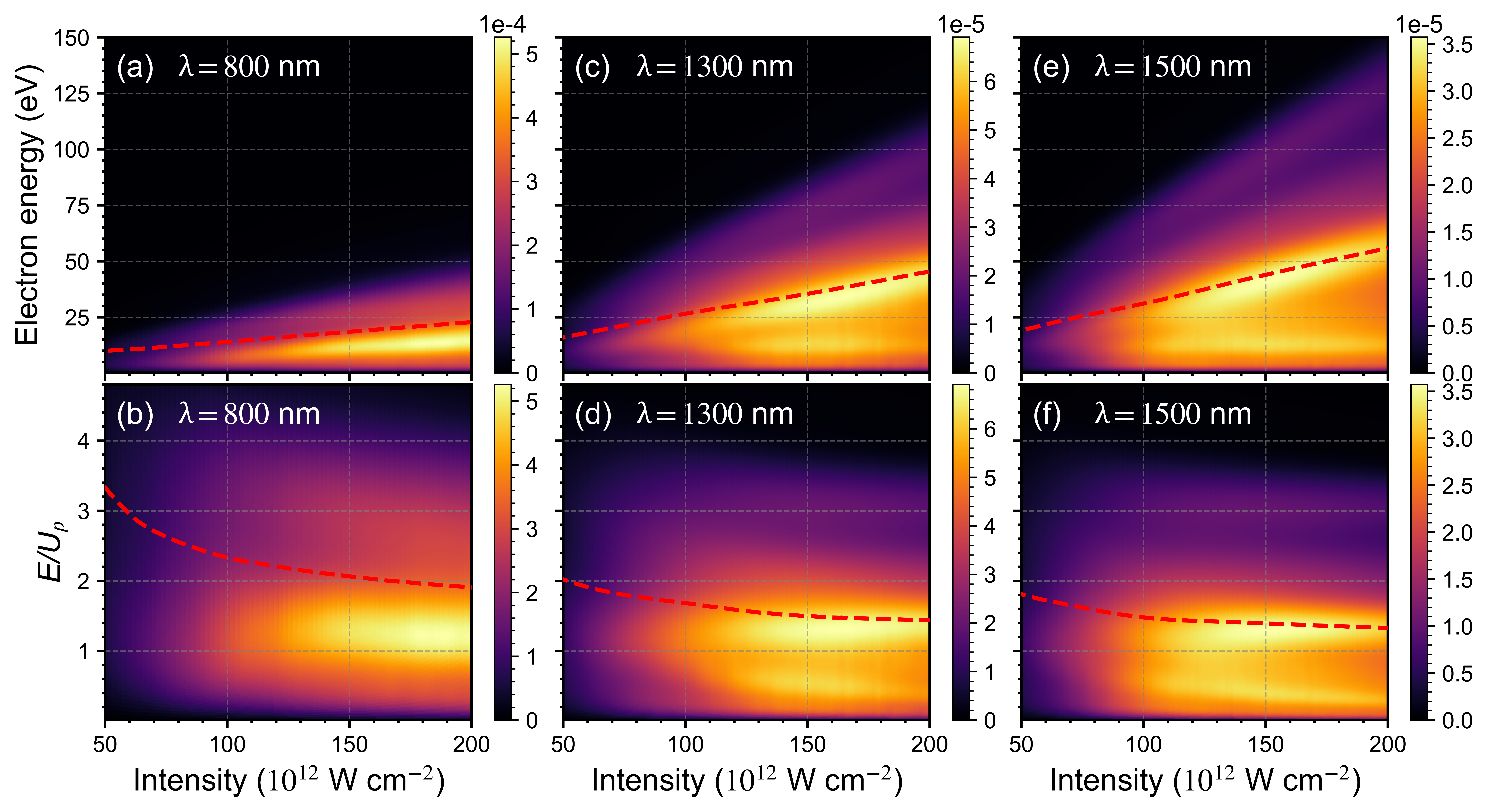}
  \caption{Calculated electron return-energy distributions as a function of laser intensity for central wavelengths of 800, 1300, and 1500 nm. Color indicates the return-electron flux. The dashed red curves indicate the mean return energy at each intensity. The bottom row presents the corresponding energies normalized to the ponderomotive energy, $U_p$. Calculations were performed by solving the time-dependent Schrödinger equation within the single-active-electron approximation.}
  \label{fig:theory}
\end{figure*}
To assess the role of rescattering in the observed wavelength dependence in Fig. \ref{fig:comparison}, Fig. \ref{fig:theory} illustrates the impact of the changing wavelength on the average return electron energies and the return electron density
for the simplest case of a single-active-electron (SAE) approximation with an atomic model potential \cite{Tong05c}, following  the procedure of Ref. \cite{Tong2007}. Transverse wavepacket spreading becomes increasingly important at longer wavelengths, reducing the probability of electron return.

The SAE model is intended to capture the dominant wavelength dependence of continuum electron propagation and return rather than molecule-specific ionization dynamics. Molecular orientation and multicenter effects can influence tunneling probabilities \cite{tong2002MTI}, but the measured ion yields are averaged over randomly oriented molecules. Multielectron excitation and channel-specific fragmentation dynamics are not explicitly included in the SAE calculation.

The top row (panels (a), (c), and (e)) shows the calculated electron return energies in eV as a function of intensity, whereas the bottom-row panels (figs. \ref{fig:theory}(b), \ref{fig:theory}(d), and \ref{fig:theory}(f)) show the calculated return energies in terms of the ponderomotive energy $U_p.$
{The vertical scales in Fig. \ref{fig:theory} show} that the return-electron flux decreases rapidly with increasing $\lambda$. As the wavelength increases from \SI{800}{nm} to \SI{1500}{nm}, the maximum electron return probability decreases by a factor of $\sim$14. 

The scaling observed from the simulations is reminiscent of the unfavorable
$\lambda^{-5}$ to $\lambda^{-6}$ single-atom scaling observed in high-harmonic generation~\cite{dimauro2007electroneff, Villenueve2009HHG, jens2012wavelengthscaling}. HHG involves coherent electron recombination of the tunneled electron, whereas excitation and fragmentation can result from inelastic electron-ion collisions. Consequently, even with a similar energy budget per electron at fixed $\gamma$, driving at shorter wavelengths (\SI{800}{nm}) yields a higher rescattering flux and hence greater fragmentation than \SI{1300}{nm} and \SI{1500}{nm}, consistent with the experimental results in Fig. \ref{fig:comparison}. However, lowering the driving frequency  also dramatically reduces laser-driven excitations inside the cation \cite{lezius2001nonadiabatic}.

To disentangle these possibilities, ion-yield measurements as a function of ellipticity were performed for several methanol fragments across the three wavelengths within the tunnel ionization regime. Since ellipticity strongly suppresses electron rescattering by laterally displacing the returning electron wavepacket from the parent ion, fragment channels driven predominantly by rescattering are expected to show a sharp reduction in yield as ellipticity deviates from zero. In contrast, channels associated with nonadiabatic excitations are expected to exhibit a comparatively weaker dependence on ellipticity. 

\begin{figure}[t]
  \centering
  \includegraphics[width=1\linewidth]{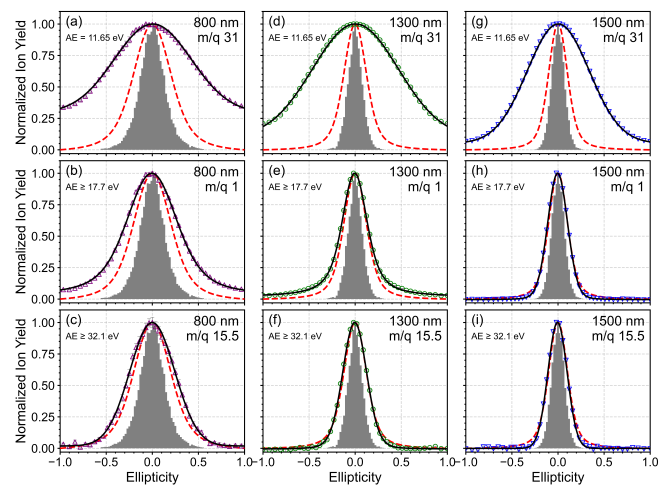}
  \caption{The top row shows ion yield as a function of ellipticity for m/q=31 (\ce{CH_3O+}), the middle row for m/q=1 (\ce{H+}), and the bottom row for m/q=15.5 (\ce{CH_3O^{2+}}). At 800, 1300, and 1500 nm, the corresponding intensities are 330 TW/cm$^2$, 200 TW/cm$^2$, and 180 TW/cm$^2$, giving Keldysh parameters $\gamma$ = 0.52, 0.41, and 0.38, respectively. Experimental ion yields and individual Monte Carlo histograms are independently normalized to unity at their maxima to facilitate comparison of ellipticity-profile shapes. The gray histograms represent simulated high-energy ($\geq 30$ eV) rescattering-event frequencies. The dashed red curves represent the simulated distributions after applying the empirical broadening determined from panel (i), uniformly across all wavelengths and channels. Gray error bars indicate the standard error of the mean of the experimental measurements. }
  \label{fig:Ellipticity}
\end{figure}

The measured ion-yield curves shown in Fig. \ref{fig:Ellipticity} reveal that the high-threshold fragment m/q=15.5 (\ce{CH2OH^{2+}}/\ce{CH3O^{2+}}) exhibits narrow Voigt-like ellipticity profiles, in contrast to the lower-appearance-energy fragment m/q=31 (\ce{CH2OH+}/\ce{CH3O+}), which exhibits broader, more Gaussian-like profiles. The m/q=15.5 curves resemble the ellipticity-dependent atomic dication yields observed in recollision-driven nonsequential double ionization \cite{dietrich1994high}. The lower bound appearance energy (AE) reported in Fig. \ref{fig:Ellipticity} for m/q=1 (\ce{H+}) is obtained from the Active Thermochemical Tables \cite{RuscicBross2020ATcT1122h}. For m/q=15.5, we use the experimentally measured appearance energy of the parent methanol dication as a lower bound on the energy required to produce this fragment, rather than as its directly measured fragmentation appearance energy \cite{Linusson2009}, while the AE for m/q=31 is taken from an experimental photoionization study \cite{Berkowitz1994}.

As the driving wavelength increases, the experimental ellipticity profiles become narrower and more closely resemble the simulated rescattering distributions, indicating a greater relative contribution from rescattering to the surviving fragmentation signal. At longer wavelengths, increased electron excursion times lead to greater transverse wavepacket spreading, reducing the absolute probability of recollision. Increasing ellipticity introduces transverse electron drift, further suppressing recollision. The wavelength dependence of these electron-return dynamics leads to an approximately inverse-wavelength scaling of the ellipticity-profile full width at half maximum (FWHM), as previously observed experimentally \cite{XU2010}. Broader experimental profiles indicate additional, less ellipticity-sensitive contributions, consistent with field-driven excitation.

The measured profiles also help assess whether a fragment arises from predominantly rescattering, inner (nonadiabatic) excitation, or their mixture.
As shown in the middle and bottom rows of Fig. \ref{fig:Ellipticity}, with increasing $\lambda$ the ellipticity dependence closely follows the recollision-associated pattern shown with the gray histograms.
Interestingly, the ellipticity dependence for the $m/q=1$ fragment at 1300 nm [Fig. 3(e)] exhibits an initial sharp decrease followed by a more gradual decay, suggesting contributions from mechanisms with different ellipticity sensitivities. At \SI{1500}{nm},  its profile closely resembles that of the dication fragment at $m/q=15.5$, consistent with a dominant rescattering contribution. In contrast, the broader response at 800 nm for $m/q=1$ (panel b) indicates a substantial contribution from a process less sensitive to ellipticity, consistent with field-driven excitation. Finally, the top row  shows that 
the formation of the $m/q=31$ fragment exhibits comparatively broad profiles, also consistent with a substantial field-driven contribution. Ellipticity profiles and corresponding FWHM values for the remaining fragments are provided in the Supplemental Material.

Monte Carlo semiclassical electron rescattering simulations were performed and confirmed the possibility of rescattering events with sufficient energy to doubly ionize methanol at various ellipticities. The potential was created using the ORCA quantum chemistry package \cite{ORCA, ORCA6}. Further 
details are described in the Supporting Information. The results of these calculations for double ionization are shown as gray-shaded distributions in Fig. 3. As expected, the ellipticity dependence becomes narrower with increasing wavelength.

Assuming that the \(m/q = 15.5\) fragment at \SI{1500}{nm} is dominated by electron rescattering, a single empirical broadening factor was determined from the experimental distribution of this channel and applied uniformly to all Monte Carlo simulation distributions. The resulting broadened profiles are shown as the dashed red curves in Fig.~\ref{fig:Ellipticity}. This procedure provides a simple first-order compensation for the approximations employed in the simulations, which otherwise overestimates the sensitivity to ellipticity.

By measuring methanol fragmentation as a function of ellipticity, wavelength, and intensity, we identify distinct signatures of electron rescattering and field-driven inner excitation. Increasing the driving wavelength reduces the probability of electron return, even when the ponderomotive energy and maximum recollision energy are held constant. The accompanying suppression of small-fragment yields, together with the largely preserved relative fragment distributions, is consistent with a reduced probability of accessing the underlying fragmentation channels rather than a significant change in their branching.

Ellipticity-dependent measurements provide an additional mechanistic discriminant. The high-threshold channels at m/q = 1 and 15.5 exhibit the strong ellipticity-dependent suppression characteristic of rescattering-driven processes, whereas the broader response of m/q = 31 indicates a substantial contribution from field-driven excitation. These results establish ellipticity-dependent mass spectrometry as a means of distinguishing competing strong-field fragmentation mechanisms in polyatomic molecules. Future measurements on molecules with more extended electronic structures, including conjugated systems, will test the generality of these findings.

This material is based upon work supported by the U.S. Department of Energy, Office of Science, Office of Basic Energy Sciences, Atomic, Molecular, and Optical Sciences Program under DE-SC0002325 to M. Dantus. XMT was supported by the Multidisciplinary Cooperative Research Program in the Center for Computational Sciences, University of Tsukuba. M. Ivanov acknowledges funding from the SFB 1477 “Light Matter Interaction at Interfaces” project number 441234705.

\bibliography{references}

@article{spanner2004reading,
  title = {Reading diffraction images in strong field ionization of diatomic molecules},
  author = {Spanner, Michael and Smirnova, Olga and Corkum, Paul B. and Ivanov, Misha Yu},
  journal = {J. Phys. B: At. Mol. Opt. Phys.},
  volume = {37},
  number = {12},
  pages = {L243--L250},
  year = {2004},
  doi = {10.1088/0953-4075/37/12/L02}
}

@article{schell2018molecular,
  title = {Molecular orbital imprint in laser-driven electron recollision},
  author = {Schell, Felix and Bredtmann, Timm and Schulz, Claus Peter and Patchkovskii, Serguei and Vrakking, Marc J. J. and Mikosch, Jochen},
  journal = {Sci. Adv.},
  volume = {4},
  number = {5},
  pages = {eaap8148},
  year = {2018},
  doi = {10.1126/sciadv.aap8148}
}

@article{smirnova2009high,
  title = {High harmonic interferometry of multi-electron dynamics in molecules},
  author = {Smirnova, Olga and Mairesse, Yann and Patchkovskii, Serguei and Dudovich, Nirit and Villeneuve, David and Corkum, Paul and Ivanov, Misha Yu},
  journal = {Nature},
  volume = {460},
  number = {7258},
  pages = {972--977},
  year = {2009},
  doi = {10.1038/nature08253}
}

@article{lein2007molecular,
  title = {Molecular imaging using recolliding electrons},
  author = {Lein, Manfred},
  journal = {J. Phys. B: At. Mol. Opt. Phys.},
  volume = {40},
  number = {16},
  pages = {R135--R173},
  year = {2007},
  doi = {10.1088/0953-4075/40/16/R01}
}

@article{lein2002interference,
  title = {Interference effects in high-order harmonic generation with molecules},
  author = {Lein, M. and Hay, N. and Marangos, J. P. and Knight, P. L.},
  journal = {Phys. Rev. A},
  volume = {66},
  number = {2},
  pages = {023805},
  year = {2002},
  doi = {10.1103/PhysRevA.66.023805}
}

@article{schafer1993above,
  title = {Above threshold ionization beyond the high harmonic cutoff},
  author = {Schafer, K. J. and Yang, Baorui and DiMauro, L. F. and Kulander, K. C.},
  journal = {Phys. Rev. Lett.},
  volume = {70},
  number = {11},
  pages = {1599--1602},
  year = {1993},
  doi = {10.1103/PhysRevLett.70.1599}
}

@article{torlina2013time,
  title = {Time-resolving electron-core dynamics during strong-field ionization in circularly polarized fields},
  author = {Torlina, Lisa and Kaushal, Jivesh and Smirnova, Olga},
  journal = {Phys. Rev. A},
  volume = {88},
  number = {5},
  pages = {053403},
  year = {2013},
  doi = {10.1103/PhysRevA.88.053403}
}

@article{uzan2020spatial,
  title = {Spatial molecular interferometry via multidimensional high-harmonic spectroscopy},
  author = {Uzan, Ayelet J. and Soifer, Hadas and Pedatzur, Oren and Clergerie, Alex and Larroque, Sylvain and Bruner, Barry D. and Pons, Bernard and Ivanov, Misha and Smirnova, Olga and Dudovich, Nirit},
  journal = {Nat. Photon.},
  volume = {14},
  number = {3},
  pages = {188--194},
  year = {2020},
  doi = {10.1038/s41566-019-0584-2}
}

@article{murray2011tunnel,
  title = {Tunnel ionization of molecules and orbital imaging},
  author = {Murray, Ryan and Spanner, Michael and Patchkovskii, Serguei and Ivanov, Misha Yu},
  journal = {Phys. Rev. Lett.},
  volume = {106},
  number = {17},
  pages = {173001},
  year = {2011},
  doi = {10.1103/PhysRevLett.106.173001}
}

@article{dietrich1994high,
  title = {High-harmonic generation and correlated two-electron multiphoton ionization with elliptically polarized light},
  author = {Dietrich, P. and Burnett, N. H. and Ivanov, M. and Corkum, P. B.},
  journal = {Phys. Rev. A},
  volume = {50},
  number = {5},
  pages = {R3585--R3588},
  year = {1994},
  doi = {10.1103/PhysRevA.50.R3585}
}

@article{dimauro2007electroneff,
  title = {Scaling of Wave-Packet Dynamics in an Intense Midinfrared Field},
  author = {Tate, J. and Auguste, T. and Muller, H. G. and Sali\`eres, P. and Agostini, P. and DiMauro, L. F.},
  journal = {Phys. Rev. Lett.},
  volume = {98},
  number = {1},
  pages = {013901},
  year = {2007},
  doi = {10.1103/PhysRevLett.98.013901}
}

@article{Tong2007,
  title = {Numerical Observation of the Rescattering Wave Packet in Laser-Atom Interactions},
  author = {Tong, X. M. and Watahiki, S. and Hino, K. and Toshima, N.},
  journal = {Phys. Rev. Lett.},
  volume = {99},
  number = {9},
  pages = {093001},
  year = {2007},
  doi = {10.1103/PhysRevLett.99.093001}
}

@article{Villenueve2009HHG,
  title = {Wavelength Scaling of High Harmonic Generation Efficiency},
  author = {Shiner, A. D. and Trallero-Herrero, C. and Kajumba, N. and Bandulet, H.-C. and Comtois, D. and L{\'e}gar{\'e}, F. and Gigu{\`e}re, M. and Kieffer, J.-C. and Corkum, P. B. and Villeneuve, D. M.},
  journal = {Phys. Rev. Lett.},
  volume = {103},
  number = {7},
  pages = {073902},
  year = {2009},
  doi = {10.1103/PhysRevLett.103.073902}
}

@article{corkum1993MPI,
  title = {Plasma perspective on strong field multiphoton ionization},
  author = {Corkum, P. B.},
  journal = {Phys. Rev. Lett.},
  volume = {71},
  number = {13},
  pages = {1994--1997},
  year = {1993},
  doi = {10.1103/PhysRevLett.71.1994}
}

@article{Blaga2012LIED,
  title = {Imaging ultrafast molecular dynamics with laser-induced electron diffraction},
  author = {Blaga, Cosmin I. and Xu, Junliang and DiChiara, Anthony D. and Sistrunk, Emily and Zhang, Kaikai and Agostini, Pierre and Miller, Terry A. and DiMauro, Louis F. and Lin, C. D.},
  journal = {Nature},
  volume = {483},
  number = {7388},
  pages = {194--197},
  year = {2012},
  doi = {10.1038/nature10820}
}

@article{lezius2002NME,
  title = {Polyatomic molecules in strong laser fields: Nonadiabatic multielectron dynamics},
  author = {Lezius, Matthias and Blanchet, Valerie and Ivanov, Misha and Stolow, Albert},
  journal = {J. Chem. Phys.},
  volume = {117},
  pages = {1575--1588},
  year = {2002},
  doi = {10.1063/1.1487823}
}

@article{Itzhak2004IDS,
  title = {Disentangling the volume effect through intensity-difference spectra: application to laser-induced dissociation of H$_2^+$},
  author = {Wang, Pengqian and Sayler, A. Max and Carnes, Kevin D. and Esry, Brett D. and Ben-Itzhak, Itzik},
  journal = {Opt. Lett.},
  volume = {30},
  number = {6},
  pages = {664--666},
  year = {2005},
  doi = {10.1364/OL.30.000664}
}

@article{Lin2010LIED,
  title = {Strong-field rescattering physics---self-imaging of a molecule by its own electrons},
  author = {Lin, C. D. and Le, Anh-Thu and Chen, Zhangjin and Morishita, Toru and Lucchese, Robert},
  journal = {J. Phys. B: At. Mol. Opt. Phys.},
  volume = {43},
  number = {12},
  pages = {122001},
  year = {2010},
  doi = {10.1088/0953-4075/43/12/122001}
}

@article{Itatani2004OrbitalImaging,
  title = {Tomographic imaging of molecular orbitals},
  author = {Itatani, J. and Levesque, J. and Zeidler, D. and Niikura, Hiromichi and P{\'e}pin, H. and Kieffer, J. C. and Corkum, P. B. and Villeneuve, D. M.},
  journal = {Nature},
  volume = {432},
  number = {7019},
  pages = {867--871},
  year = {2004},
  doi = {10.1038/nature03183}
}

@article{Niikura2003,
  title = {Probing molecular dynamics with attosecond resolution using correlated wave packet pairs},
  author = {Niikura, Hiromichi and L{\'e}gar{\'e}, F. and Hasbani, R. and Ivanov, Misha Yu and Villeneuve, D. M. and Corkum, P. B.},
  journal = {Nature},
  volume = {421},
  number = {6925},
  pages = {826--829},
  year = {2003},
  doi = {10.1038/nature01430}
}

@article{corkum1995EnhancedIonization,
  title = {Role of Electron Localization in Intense-Field Molecular Ionization},
  author = {Seideman, Tamar and Ivanov, M. Yu. and Corkum, P. B.},
  journal = {Phys. Rev. Lett.},
  volume = {75},
  number = {15},
  pages = {2819--2822},
  year = {1995},
  doi = {10.1103/PhysRevLett.75.2819}
}

@article{tong2002MTI,
  title = {Theory of molecular tunneling ionization},
  author = {Tong, X. M. and Zhao, Z. X. and Lin, C. D.},
  journal = {Phys. Rev. A},
  volume = {66},
  number = {3},
  pages = {033402},
  year = {2002},
  doi = {10.1103/PhysRevA.66.033402}
}

@article{Suchowski2008starkshifts,
  title = {Spatio-temporal coherent control of atomic systems: weak to strong field transition and breaking of symmetry in 2D maps},
  author = {Suchowski, H. and Natan, A. and Bruner, B. D. and Silberberg, Y.},
  journal = {J. Phys. B: At. Mol. Opt. Phys.},
  volume = {41},
  number = {7},
  pages = {074008},
  year = {2008},
  doi = {10.1088/0953-4075/41/7/074008}
}

@article{Liu2007starkshifts,
  title = {Dynamics of multilevel molecules and pulse propagation beyond rotating wave approximation near two-photon resonance},
  author = {Liu, Ji-Cai and Wang, Chuan-Kui and Gel'mukhanov, Faris},
  journal = {Phys. Rev. A},
  volume = {76},
  number = {4},
  pages = {043422},
  year = {2007},
  doi = {10.1103/PhysRevA.76.043422}
}

@article{Wu2012Enhancedionization,
  title = {Probing the tunnelling site of electrons in strong field enhanced ionization of molecules},
  author = {Wu, J. and Meckel, M. and Schmidt, L. Ph. H. and Kunitski, M. and Voss, S. and Sann, H. and Kim, H. and Jahnke, T. and Czasch, A. and D{\"o}rner, R.},
  journal = {Nat. Commun.},
  volume = {3},
  number = {1},
  pages = {1113},
  year = {2012},
  doi = {10.1038/ncomms2130}
}

@article{corkum1996Enhancedionization,
  title = {Observation of Enhanced Ionization of Molecular Ions in Intense Laser Fields},
  author = {Constant, E. and Stapelfeldt, H. and Corkum, P. B.},
  journal = {Phys. Rev. Lett.},
  volume = {76},
  number = {22},
  pages = {4140--4143},
  year = {1996},
  doi = {10.1103/PhysRevLett.76.4140}
}

@article{jens2012wavelengthscaling,
  title = {Strong-field approximation for the wavelength scaling of high-harmonic generation},
  author = {Austin, Dane R. and Biegert, Jens},
  journal = {Phys. Rev. A},
  volume = {86},
  number = {2},
  pages = {023813},
  year = {2012},
  doi = {10.1103/PhysRevA.86.023813}
}

@article{Rajgara2003RescatteringFrag,
  title = {Electron rescattering and the fragmentation dynamics of molecules},
  author = {Rajgara, F. A. and Safaei, R. and Chin, S. L.},
  journal = {Phys. Rev. A},
  volume = {68},
  number = {2},
  pages = {023407},
  year = {2003},
  doi = {10.1103/PhysRevA.68.023407}
}

@article{keldysh1964ionization,
  title = {Ionization in the field of a strong electromagnetic wave},
  author = {Keldysh, L. V.},
  journal = {J. Exp. Theor. Phys.},
  volume = {20},
  pages = {1307--1314},
  year = {1965}
}

@article{nubbemeyer2008strong,
  title = {Strong-field tunneling without ionization},
  author = {Nubbemeyer, T. and G{\"o}rling, K. and Saenz, A. and Eichmann, U. and Sandner, W.},
  journal = {Phys. Rev. Lett.},
  volume = {101},
  number = {23},
  pages = {233001},
  year = {2008},
  doi = {10.1103/PhysRevLett.101.233001}
}

@article{mavsin2018electron,
  title = {Electron correlations and pre-collision in the re-collision picture of high harmonic generation},
  author = {Ma{\v{s}}{\'i}n, Zden{\v{e}}k and Harvey, Alex G. and Spanner, Michael and Patchkovskii, Serguei and Ivanov, Misha and Smirnova, Olga},
  journal = {J. Phys. B: At. Mol. Opt. Phys.},
  volume = {51},
  number = {13},
  pages = {134006},
  year = {2018},
  doi = {10.1088/1361-6455/aac598}
}

@article{guo1998single,
  title = {Single and double ionization of diatomic molecules in strong laser fields},
  author = {Guo, Chunlei and Li, Ming and Nibarger, John P. and Gibson, George N.},
  journal = {Phys. Rev. A},
  volume = {58},
  number = {6},
  pages = {R4271--R4274},
  year = {1998},
  doi = {10.1103/PhysRevA.58.R4271}
}

@article{ORCA,
  title = {The ORCA program system},
  author = {Neese, F.},
  journal = {Wiley Interdiscip. Rev. Comput. Mol. Sci.},
  volume = {2},
  number = {1},
  pages = {73--78},
  year = {2012},
  doi = {10.1002/wcms.81}
}

@article{ORCA6,
  title = {Software Update: The ORCA Program System---Version 6.0},
  author = {Neese, Frank},
  journal = {Wiley Interdiscip. Rev. Comput. Mol. Sci.},
  volume = {15},
  number = {2},
  pages = {e70019},
  year = {2025},
  doi = {10.1002/wcms.70019}
}

@article{Shusterman2022,
  title = {Coulomb Explosion Dynamics of Multiply Charged para-Nitrotoluene Cations},
  author = {Shusterman, Jacob M. and Gutsev, Gennady L. and L{\'o}pez Pe{\~n}a, Hugo A. and Ramachandran, B. Ramu and Tibbetts, Katharine Moore},
  journal = {J. Phys. Chem. A},
  volume = {126},
  number = {38},
  pages = {6617--6627},
  year = {2022},
  doi = {10.1021/acs.jpca.2c04395}
}

@article{bohinski2013measurement,
  title = {Measurement of an electronic resonance in a ground-state, gas-phase acetophenone cation via strong-field mass spectrometry},
  author = {Bohinski, Timothy and Moore Tibbetts, Katharine and Tarazkar, Maryam and Romanov, Dmitri and Matsika, Spiridoula and Levis, Robert J.},
  journal = {J. Phys. Chem. Lett.},
  volume = {4},
  number = {10},
  pages = {1587--1591},
  year = {2013},
  doi = {10.1021/jz400516h}
}

@article{posthumus2004dynamics,
  title = {The dynamics of small molecules in intense laser fields},
  author = {Posthumus, J. H.},
  journal = {Rep. Prog. Phys.},
  volume = {67},
  number = {5},
  pages = {623--665},
  year = {2004},
  doi = {10.1088/0034-4885/67/5/R01}
}

@article{lezius2001nonadiabatic,
  title = {Nonadiabatic multielectron dynamics in strong field molecular ionization},
  author = {Lezius, M. and Blanchet, Val{\'e}rie and Rayner, D. M. and Villeneuve, D. M. and Stolow, Albert and Ivanov, Misha Yu},
  journal = {Phys. Rev. Lett.},
  volume = {86},
  number = {1},
  pages = {51--54},
  year = {2001},
  doi = {10.1103/PhysRevLett.86.51}
}

@article{shchatsinin2009multiphoton,
  title = {Multiphoton excitation and ionization by elliptically polarized, intense short laser pulses: Recognizing multielectron dynamics and doorway states in C$_{60}$ vs Xe},
  author = {Shchatsinin, I. and Ritze, H.-H. and Schulz, C. P. and Hertel, I. V.},
  journal = {Phys. Rev. A},
  volume = {79},
  number = {5},
  pages = {053414},
  year = {2009},
  doi = {10.1103/PhysRevA.79.053414}
}

@article{bhardwaj2004recollision,
  title = {Recollision during the high laser intensity ionization of C$_{60}$},
  author = {Bhardwaj, V. R. and Corkum, P. B. and Rayner, D. M.},
  journal = {Phys. Rev. Lett.},
  volume = {93},
  number = {4},
  pages = {043001},
  year = {2004},
  doi = {10.1103/PhysRevLett.93.043001}
}

@article{Popruzhenko_2014,
  title = {Keldysh theory of strong field ionization: history, applications, difficulties and perspectives},
  author = {Popruzhenko, S. V.},
  journal = {J. Phys. B: At. Mol. Opt. Phys.},
  volume = {47},
  number = {20},
  pages = {204001},
  year = {2014},
  doi = {10.1088/0953-4075/47/20/204001}
}

@article{Dantus_science2024,
  title = {Ultrafast studies of elusive chemical reactions in the gas phase},
  author = {Dantus, Marcos},
  journal = {Science},
  volume = {385},
  number = {6709},
  pages = {eadk1833},
  year = {2024},
  doi = {10.1126/science.adk1833}
}

@article{Konar2014,
  title = {Polyatomic Molecules under Intense Femtosecond Laser Irradiation},
  author = {Konar, Arkaprabha and Shu, Yinan and Lozovoy, Vadim V. and Jackson, James E. and Levine, Benjamin G. and Dantus, Marcos},
  journal = {J. Phys. Chem. A},
  volume = {118},
  number = {49},
  pages = {11433--11450},
  year = {2014},
  doi = {10.1021/jp505498t}
}

@article{Berkowitz1994,
  title = {Three methods to measure RH bond energies},
  author = {Berkowitz, Joseph and Ellison, G. Barney and Gutman, David},
  journal = {J. Phys. Chem.},
  volume = {98},
  number = {11},
  pages = {2744--2765},
  year = {1994},
  doi = {10.1021/j100062a009}
}

@misc{RuscicBross2020ATcT1122h,
  author = {Ruscic, Branko and Bross, David H.},
  title = {{Active Thermochemical Tables (ATcT) values based on ver. 1.122h of the Thermochemical Network}},
  year = {2020}
}

@article{Tong05c,
  title = {Empirical formula for static field ionization rates of atoms and molecules by lasers in the barrier-suppression regime},
  author = {Tong, X. M. and Lin, C. D.},
  journal = {J. Phys. B: At. Mol. Opt. Phys.},
  volume = {38},
  number = {15},
  pages = {2593--2600},
  year = {2005},
  doi = {10.1088/0953-4075/38/15/001}
}

@article{Wang2017,
  title = {Wavelength scaling of atomic nonsequential double ionization in intense laser fields},
  author = {Wang, YanLan and Xu, SongPo and Chen, YongJu and Kang, HuiPeng and Lai, XuanYang and Quan, Wei and Liu, XiaoJun and Hao, XiaoLei and Li, WeiDong and Hu, ShiLin and Chen, Jing and Becker, Wilhelm and Chu, Wei and Yao, Jinping and Zeng, Bin and Cheng, Ya and Xu, ZhiZhan},
  journal = {Phys. Rev. A},
  volume = {95},
  number = {6},
  pages = {063415},
  year = {2017},
  doi = {10.1103/PhysRevA.95.063415}
}

@article{Tanaka2009,
  title = {Ionization and Fragmentation of Alkylphenols by 0.8--1.5 $\mu$m Femtosecond Laser Pulses},
  author = {Tanaka, Michinori and Kawaji, Mariko and Yatsuhashi, Tomoyuki and Nakashima, Nobuaki},
  journal = {J. Phys. Chem. A},
  volume = {113},
  number = {44},
  pages = {12056--12062},
  year = {2009},
  doi = {10.1021/jp902557v}
}

@article{BOHINSKI2014,
title = {Radical cation spectroscopy of substituted alkyl phenyl ketones via tunnel ionization},
journal = {Chemical Physics},
volume = {442},
pages = {81-85},
year = {2014},
note = {Papers from the 11th International Conference on Femtochemistry (FEMTO11) that took place at The Technical University of Denmark (DTU), Copenhagen},
issn = {0301-0104},
doi = {https://doi.org/10.1016/j.chemphys.2014.05.019},
url = {https://www.sciencedirect.com/science/article/pii/S0301010414001591},
author = {Timothy Bohinski and Katharine {Moore Tibbetts} and Kristin Munkerup and Maryam Tarazkar and Dmitri A. Romanov and Spiridoula Matsika and Robert J. Levis}
}

@Article{Li2021,
author={Li, Yuanyue
and Kind, Tobias
and Folz, Jacob
and Vaniya, Arpana
and Mehta, Sajjan Singh
and Fiehn, Oliver},
title={Spectral entropy outperforms MS/MS dot product similarity for small-molecule compound identification},
journal={Nature Methods},
year={2021},
month={Dec},
day={01},
volume={18},
number={12},
pages={1524-1531},
issn={1548-7105},
doi={10.1038/s41592-021-01331-z},
url={https://doi.org/10.1038/s41592-021-01331-z}
}

@article{Linusson2009,
  title = {Double photoionization of alcohol molecules},
  author = {Linusson, P. and Stenrup, M. and Larson, {\AA{}}. and Andersson, E. and Heijkenskj{\"o}ld, F. and Andersson, P. and Eland, J. H. D. and Karlsson, L. and Rubensson, J.-E. and Feifel, R.},
  journal = {Phys. Rev. A},
  volume = {80},
  issue = {3},
  pages = {032516},
  numpages = {6},
  year = {2009},
  month = {Sep},
  publisher = {American Physical Society},
  doi = {10.1103/PhysRevA.80.032516},
  url = {https://link.aps.org/doi/10.1103/PhysRevA.80.032516}
}

@article{Xu2010,
author = {Han Xu and Hui Xiong and Bin Zeng and Wei Chu and Yuxi Fu and Jingpin Yao and Jing Chen and Xiaojun Liu and Ya Cheng and Zhizhan Xu},
journal = {Opt. Lett.},
number = {4},
pages = {472--474},
publisher = {Optica Publishing Group},
title = {Wavelength scaling of elliptical-polarization dependence of high-order harmonic generation},
volume = {35},
month = {Feb},
year = {2010},
url = {https://opg.optica.org/ol/abstract.cfm?URI=ol-35-4-472},
doi = {10.1364/OL.35.000472}
}

\end{document}